\documentclass[draft]{agujournal}
\usepackage{epstopdf}

\usepackage{graphicx}
\usepackage{amssymb,amsmath, amssymb,amsbsy, bm}
\usepackage{wasysym}

\begin{document}

\title{Exact gravity field for polyhedrons with  polynomial density contrasts of arbitrary orders}

\authors{ Zhengyong Ren\affil{1}, Chaojian Chen\affil{1,2,*}, Yiyuan Zhong\affil{1}, Huang Chen\affil{1}, \\ Thomas Kalscheuer\affil{3}, Hansruedi Maurer\affil{2}, Jingtian Tang\affil{1*},  Xiangyun Hu\affil{4}}

\affiliation{1}{School of Geosciences and Info-Physics, Central South University, Changsha, Hunan, China.}
\affiliation{2}{Department of Earth Sciences, Institute of Geophysics, ETH Zurich, Zurich, Switzerland.}
\affiliation{3}{Department of Earth Sciences, Uppsala University, Uppsala, Sweden.}
\affiliation{4}{Institute of Geophysics and Geomaties, Chinese University of Geoscience, Wuhan, Hubei, China.}

\correspondingauthor{Chaojian Chen}{chenchaojian@csu.edu.cn}
\correspondingauthor{Jingtian Tang  }{jttang@csu.edu.cn}

\begin{keypoints}
\item Anomaly mass target is represented by  a set of  polyhedral mass elements
\item Density contrast is  a polynomial function of arbitrary orders which can simultaneously vary in both horizontal and vertical directions
\item Exact solutions of gravity fields are singularity-free
\end{keypoints}

\begin{abstract}
Computing gravity field of a  mass body is a core routine to image anomalous density structures in the Earth.  In this study, we report  the existence of analytical routines to accurately compute the gravity potential and gravity field of  a general polyhedral mass body. The density contrasts in the polyhedral body can be general polynomial functions up to arbitrary non-negative orders and also can vary in both horizontal and vertical directions. The newly derived analytical expressions of gravity fields are also singularity-free which means that observation sites can have arbitrary geometric relationships with polyhedral mass bodies.  One synthetic prismatic body with different density contrasts is used to verify  the accuracies of our new closed-form solutions. Excellent agreements are obtained among our solutions and other   published solutions. Our work  is  the first-of-its kind to completely answer the fundamental question on existence of analytic solutions of gravitational field for general mass bodies. It may put an end of  searching closed-form solutions in gravity surveying.
\end{abstract}

\section{Introduction}
Gravity data sets can be measured by gravimeters installed at land-based stations, on airplanes and ships, and even on satellites. Due to rapid developments of high precise gravimeters and financial invests, a large amount of gravity data sets are available to explore the minerals in the shallow surfaces, to investigate regional structures in the upper crust, even to image the density distributions in the mantle \citep{Hautmann2013grl, Panet2014Mapping, Ye2016gji,Cericia2013geophysics}.  To effectively extract anomalous density structures from measured gravity data sets,  there is an urgent demand to develop routines aiming to  forward compute accurately gravity fields for a given mass body \citep{Barnett1976, HofmannMoritz2006,ren2017JGR}.

The analytical expression (or closed-form solution)  for a polyhedral mass body is the routine of choice as it not only can well approximates complicated mass bodies in the Earth, but also  offers ultimate accuracies. Since 1950s, intensive studies have been devoted to search closed-form formulae of gravity field for a  polyhedral mass body with polynomial density contrasts as polynomial density contrasts can easily approximate the complicated mass density contributions in the Earth. For low order polynomial cases, a large amount of closed-form solutions were derived, such as the constant case \citep{Paul1974, Barnett1976, Okabe1979, Holstein1996, holstein1999comparison, Tsoulis01032001, holstein2002gravimagnetic, Tsoulis2012}, the linear case \citep{wilton1984potential, Holstein2003, Hamayun2009, DUrso2014, Ren2017SG}, as well as the quadratic and cubic cases \citep{DUrso2017, Ren2018SG, Ren2018Geophysics}. However,  density contrasts in the realistic Earth generally have more complicated distributions than those described by low order polynomials. For instance, exogenetic and endogenetic processes in the earth generally can change the mass density structures of the crust and mantle into three dimensional structures whose densities can vary both in horizontal and vertical directions \citep{Martin-Atienza01121999}. Therefore, it is more reasonable to consider a general density contrast of polynomial forms (of arbitrary orders) to more better approximate the true complicated density contrasts of the Earth.

In very recent years, several efforts have been paid to address above issues. For a  prism,  \cite{Jiang2017geophysics}, \cite{Roland2018} and \cite{Fukushima2018gji}
successfully find   closed-form solutions of gravity field for density contrast  varying along depth  by following  a polynomial function of  arbitrary orders.  For a prism,  \cite{Zhang2018gji} derived   closed-form solutions of gravity field for
density contrast varying both in horizontal and vertical directions and following polynomial functions of arbitrary orders.  For a polyhedral prism,  \cite{chen2018gji} derived closed-form solutions of vertical gravity field  with vertical polynomial density contrast up to arbitrary orders. A logic and theoretically important step is to find   closed-form solutions of gravity field for  a polyhedral body with arbitrary order polynomial density contrasts, which has not been addressed yet so far.

This study  reports our latest finding which has successfully addressed above step.  We find analytic formulae of both gravity potential and gravity field for a polyhedral body with polynomial density contrast up to arbitrary orders. The density contrasts in the polyhedral body can simultaneously vary  in both horizontal and vertical directions. We use two synthetic models to verify accuracies of the new analytic formulae.

\section{Theory and new analytic formulae}
Let us adopt a right-handed Cartesian coordinate system where $x$-axis, $y$-axis are horizontally directed, and positive $z$-axis is vertically downward. For a  polyhedral mass body $H$,   gravity potential $(\phi)$ and gravity field $(\mathbf{g})$ at an observation site $\mathbf{r}^{\prime}$ are  \citep{blakely1996}:
\begin{eqnarray}
\phi(\mathbf{r'})&=&G\iiint_{H}\frac{\lambda(\mathbf{r})}{R}dv, \label{gravity_potential}\\
\mathbf{g}(\mathbf{r'})&=&\nabla_{\mathbf{r'}}\phi(\mathbf{r'})=G\iiint_{H}\lambda(\mathbf{r})\nabla_{\mathbf{r'}}\frac{1}{R}dv, \label{gravity_field}
\label{1}
\end{eqnarray}
where $G=6.673\times10^{-11}\,\text{m}^3\text{kg}^{-1}\text{s}^{-2}$ is the gravitational  constant, and $R=|\mathbf{r}-\mathbf{r'}|$ is the distance from the observation site $\mathbf{r'}$ to a running integral point $\mathbf{r}$ in   body $H$, with $\mathbf{r}\in H$. The  density contrast in the polyhedral body is denoted by $\mathbf{\lambda(r)}$, which is a polynomial function varying in both  horizontal and vertical directions:
\begin{eqnarray}
\lambda(\mathbf{r})=\sum_{p=0}^{n}\sum_{q=0}^{n-p}\sum_{t=0}^{n-(p+q)}a_{pqt}x^{p}y^{q}z^{t},
\label{density}
\end{eqnarray}
where $a_{pqt}$ is the constant coefficient which can be estimated by fitting  measured field data sets, etc, from borehole \citep{blakely1996}. Non-negative integer $n$ is the order of the polynomial function ($n\geq0$).  Non-negative integers $p,q,t$ are three suborders along  $x$-, $y$- and $z$-directions, respectively ($0\leq p+q+t\leq n$).

Seeking for simplicity, we first move the observation site $\mathbf{r}^{\prime}$ to be the origin of the Cartesian coordinate system, that is $\mathbf{r}^{\prime}=(0,0,0)$.  After repeatedly using  integration by parts, we finally obtain  following formulae for both $\phi$ and $\mathbf{g}$:
\begin{eqnarray}
\phi(\mathbf{r'})&=&G\sum_{p=0}^{n}\sum_{q=0}^{n-p}\sum_{t=0}^{n-(p+q)}a_{pqt}\phi(p,q,t),\\
\mathbf{g}(\mathbf{r'})&=&-G\sum_{p=0}^{n}\sum_{q=0}^{n-p}\sum_{t=0}^{n-(p+q)}a_{pqt}\mathbf{g}(p,q,t),
\end{eqnarray}
where
\begin{eqnarray}
\phi(p,q,t)&=&
\left\{ {\begin{array}{*{20}{c}}
\sum_{i=1}^{N}(\hat{\mathbf{n}}_i\cdot\hat{\mathbf{x}}) I_s(p-1,q,t,1)-(p-1)\cdot I_v(p-2,q,t,1),\quad p\geq2,\\
\sum_{i=1}^{N}(\hat{\mathbf{n}}_i\cdot\hat{\mathbf{x}}) I_s(0,q,t,1),\quad p=1,\\
I_v(0,q,t,-1),\quad p=0,
\end{array}} \right.,\label{final_gravity_potential}\\
{\mathbf{g}}(p,q,t) &=& \sum_{i=1}^{N}\hat{\mathbf{n}}_{i} I_s(p,q,t,-1)
-p\cdot\phi(p-1,q,t)\hat{\mathbf{x}}
-q\cdot\phi(p,q-1,t)\hat{\mathbf{y}}
-t\cdot\phi(p,q,t-1)\hat{\mathbf{z}},\nonumber\\
\label{final_gravity_field}
\end{eqnarray}
where $N$ is the number of polygonal surfaces of the polyhedral body $H$, and $\hat{\mathbf{n}}_{i}$ is the outward normal vector of the $i$-th polygonal facet $\partial H_i$. The unit vectors along $x$-, $y$-, and $z$-directions are denoted by $\hat{\mathbf{x}}$, $\hat{\mathbf{y}}$, $\hat{\mathbf{z}}$, respectively. Symbols $I_v$ and $I_s$ represent the volume integrals and surface integrals, $I_v(a,b,c,w) = \iiint_{H} x^a y^b z^cR^wdv$, $I_s(\alpha,\beta,\gamma,\delta)=\iint_{\partial H_i}x^\alpha y^\beta z^\gamma R^{\delta}ds$, respectively, where $a,b,c,w,\alpha,\beta,\gamma,\delta$ are integer indices. Analytical expressions for $I_v$ and $I_s$ are listed in Text S1 of the supporting information.

\clearpage
\begin{table}
\centering
\caption{Comparison of our formula to other closed-form solutions of a  polyhedral body for gravity potential $(\phi)$, and/or gravity field $(\mathbf{g})$. Symbol $-$ represents  singularity, and symbol $\surd$ indicates   singularity-free.}
\begin{tabular}{llll}
\hline
 Maxinium order & Singularity & Components & References \\
       &  free &   &   \\
\hline
 Constant        &$-$     &$g_z$                          &\cite{Paul1974,Barnett1976}   \\
 Constant        &$-$     &$\mathbf{g}$        &\cite{Okabe1979}   \\
 Constant        &$\surd$ &$\mathbf{g}$                   &\cite{Pohanka1988, Holstein1996}   \\
                 &        &                               &\cite{Hansen1999, holstein1999comparison}\\
                 &        &                               &\cite{Tsoulis01032001, Conway2015}\\
 Constant        &$\surd$ &$\phi,\mathbf{g}$              &\cite{holstein2002gravimagnetic}   \\
                 &        &       &\cite{Petrovic1996,Tsoulis2012}   \\
                 &        &       &\cite{DUrso2013, DUrso2014}   \\
 Linear          &$\surd$ &$\phi$                         &\cite{Hamayun2009}   \\
 Linear          &$\surd$ &$\mathbf{g}$                   &\cite{Pohanka1998, Hansen1999}   \\
 Linear          &$\surd$ &$\phi,\mathbf{g}$              &\cite{Ren2017SG}   \\
                 &        &    &\cite{Holstein2003, DUrso2014linear}   \\
 Cubic order     &$\surd$ &$\phi,\mathbf{g}$        &\cite{DUrso2017,Ren2018Geophysics}  \\
Arbitrary order &$\surd$ &$\phi,\mathbf{g}$        &This study     \\
 \hline
\end{tabular}
\label{table1}
\end{table}

During last 44 years, intensive studies were conducted to derive the closed-form solutions of gravity field for a polyhedral body. For reader's convenience, we have listed these solutions in Table (\ref{table1}). Seen from this table,  it is clear that our work unifies previous results.

\section{Accuracy validation}
\begin{figure}
\centering
\includegraphics[width=12cm]{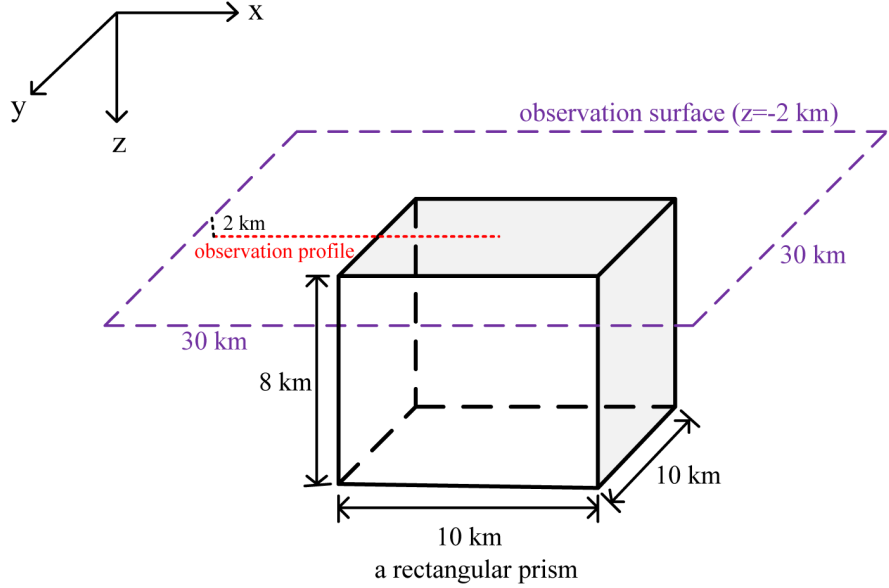}
\caption{Illustration of the tested rectangular prism. Dashed line in red color represents the observation site which lies on the surface of the prism. The purple line represents the horizontal observation surface with $z=-2\,\text{km}$.}
\label{testmodelprism}
\end{figure}

A uniform prismatic body (see Figure \ref{testmodelprism}) with density contrast of $\lambda(\mathbf{r})= 2670\,\text{kg/m}^3$ is tested to verify the accuracies of our closed-form solutions. This model is taken from previous study  \citep{garciacubic2005geophysics} which has a size of $x=[10\,\text{km},20\,\text{km}]$, $y=[10\,\text{km},20\,\text{km}]$ and $z=[0\,\text{km},8\,\text{km}]$. Sixteen observations sites are equally spaced along a profile from $x=0\,\text{km}$ to $x=15\,\text{km}$, with $y=15\,\text{km}$ and $z=0\,\text{km}$. The gravity potentials ($\phi$) and vertical gravity fields ($g_z$)  are calculated by four different methods, which are \cite{nagy2000jog}'s analytical method (singularity-free formulae derived for a homogeneous  prismatic body), \cite{Tsoulis2012}'s  analytical method (singularity-free formulae designed for a  polyhedral body), high-order Gaussian numerical quadrature method with $512\times512\times512$ quadrature points \citep{Davis1984Methods} (working for a  polyhedral body but with singularity), and our new formulae  (singularity-free formulae designed for a  polyhedral body).

Tables (\ref{t2}) and (\ref{t3}) show the gravity potentials and the vertical gravity fields computed by these four  methods. When  observation sites are outside the rectangular prism with  $x$-coordinates ranging from $0\,\text{km}$ to $9\,\text{km}$ where no singularities exist,   four   solutions agree well with each other. The solutions of our   and \cite{nagy2000jog}'s formulae are identical up to the 13th significant digit, which are at the 7th significant digit between our formula and \cite{Tsoulis2012}'s formula, and 11th significant digit between our formula and the high-order Gaussian quadrature rule. When  observation sites locate on   edge or  surface of the prism with   $x$-coordinates ranging from $10\,\text{km}$ to $15\,\text{km}$ where  mathematic singularities exist, the high-order Gaussian quadrature with 134,217,728 quadrature points cannot produce desired gravity fields. The relative errors between our solutions and high-order Gaussian quadrature rule's solutions are approximately $10^{-2}\%$ for gravity potential and $10^{-1}\%$ for vertical gravity field, respectively.  Solutions of these three singularity-free analytical formulae still agree well with each other.  Relative errors of our formulae referring to \cite{nagy2000jog}'s formulae  are less than $10^{-13}\%$ both for gravity potential and vertical gravity field. The relative errors are about $10^{-7}\%$ by referring to  \cite{Tsoulis2012}'s  solutions,  which suggests our solutions and \cite{nagy2000jog}'s solutions maybe more accurate than \cite{Tsoulis2012}'s  solutions. 

\begin{sidewaystable}
\centering
\caption{Comparison of computed gravity potential ($\phi$)  by our new formula, \cite{nagy2000jog}'s formula,  \cite{Tsoulis2012}'s formula and the high-order Gauss quadrature rule \citep{Davis1984Methods}. The observation sites are equally spaced on the profile from $x=0\,\text{km}$ to $x=15\,\text{km}$ with $y=15\,\text{km}$ and $z=0\,\text{km}$. Differences of later three solutions referring to our closed-form solutions are marked by bold face.}
\begin{tabular}{crrrr}
\hline
 & \multicolumn{4}{c}{$\phi$ ($\text{m}^2\,\text{s}^{-2}$) on the observation profile ($y=15\,\text{km}$ and $z=0\,\text{m}$)}\\
$x (\,\text{km})$  & Our  solution &  \cite{nagy2000jog}'s solution &\cite{Tsoulis2012}'s solution &  Gauss quadrature rule's solution\\\hline
0   & 9.213370778767388E+0  & 9.2133707787673\textbf{96}E+0 & 9.2133\textbf{69100056041}E+0  &9.2133707787663\textbf{66}E+0\\
1   & 9.824631495073580E+0  & 9.8246314950735\textbf{76}E+0 & 9.82463\textbf{0015757330}E+0 &9.82463149507\textbf{4603}E+0\\
2   & 1.051797364599913E+1  & 1.0517973645999\textbf{04}E+1 & 1.051797\textbf{237193737}E+1 &1.051797364\textbf{600060}E+1\\
3   & 1.130972911860967E+1  & 1.1309729118609\textbf{65}E+1 & 1.130972\textbf{805682759}E+1 &1.13097291186\textbf{1287}E+1\\
4   & 1.222032550160522E+1  & 1.2220325501605\textbf{19}E+1 & 1.222032\textbf{466058624}E+1 &1.222032550160\textbf{315}E+1\\
5   & 1.327538330028535E+1  & 1.3275383300285\textbf{37}E+1 & 1.327538\textbf{269034185}E+1 &1.327538330028\textbf{375}E+1\\
6   & 1.450701005685137E+1  & 1.4507010056851\textbf{29}E+1 & 1.45070\textbf{0969058767}E+1 &1.450701005685\textbf{489}E+1\\
7   & 1.595517547999473E+1  & 1.5955175479994\textbf{65}E+1 & 1.5955175\textbf{37286842}E+1 &1.595517547999\textbf{217}E+1\\
8   & 1.766884275902838E+1  & 1.7668842759028\textbf{33}E+1 & 1.7668842\textbf{92999224}E+1 &1.766884275902\textbf{748}E+1\\
9   & 1.970619558083274E+1  & 1.9706195580832\textbf{89}E+1 & 1.970619\textbf{605299300}E+1 &1.970619558083\textbf{330}E+1\\
10  & 2.213300296314875E+1  & 2.2133002963148\textbf{86}E+1 & 2.213300\textbf{307117277}E+1 &2.213\textbf{298105404876}E+1\\
11  & 2.445867695151962E+1  & 2.4458676951519\textbf{59}E+1 & 2.4458676\textbf{67261200}E+1 &2.44586\textbf{3325909497}E+1\\
12  & 2.618973014389963E+1  & 2.6189730143899\textbf{57}E+1 & 2.6189730\textbf{09759514}E+1 &2.6189\textbf{68115907312}E+1\\
13  & 2.738306438467226E+1  & 2.7383064384672\textbf{19}E+1 & 2.7383064\textbf{50143388}E+1 &2.73830\textbf{1380637997}E+1\\
14  & 2.808152004294769E+1  & 2.808152004294769E+1          & 2.8081520\textbf{25629653}E+1 &2.8081\textbf{46845667066}E+1\\
15  & 2.831139474536049E+1  & 2.83113947453604\textbf{8}E+1 & 2.8311394\textbf{99069494}E+1 &2.83113\textbf{3673809870}E+1\\
\hline
\end{tabular}
\label{t2}
\end{sidewaystable}

\begin{sidewaystable}
\centering
\caption{Similar to Table \ref{t2}, but for the vertical gravity field ($g_z$).}
\begin{tabular}{crrrr}
\hline
 & \multicolumn{3}{c}{$g_z$ ($\text{m}\,\text{s}^{-2}$) on the observation profile ($y=15\,\text{km}$ and $z=0\,\text{m}$)}\\
$x (\,\text{km})$  & Our  solution &  \cite{nagy2000jog}'s solution &\cite{Tsoulis2012}'s solution &  Gauss quadrature rule's solution\\\hline
0   & 1.569077805220523E-4  & 1.569077805220\textbf{415}E-4 & 1.569077\textbf{9172288047}E-4  &1.569077805220\textbf{810}E-4\\
1   & 1.903530880423478E-4  & 1.903530880423\textbf{529}E-4 & 1.90353\textbf{10163066231}E-4 &1.903530880423\textbf{669}E-4\\
2   & 2.336177366177579E-4  & 2.336177366177\textbf{611}E-4 & 2.336177\textbf{5329451025}E-4 &2.336177366177\textbf{468}E-4\\
3   & 2.903939376743299E-4  & 2.9039393767432\textbf{63}E-4 & 2.903939\textbf{5840403946}E-4 &2.90393937674\textbf{2514}E-4\\
4   & 3.660760399102218E-4  & 3.6607603991022\textbf{50}E-4 & 3.660760\textbf{6604248237}E-4 &3.6607603991022\textbf{69}E-4\\
5   & 4.687144103570855E-4  & 4.68714410357085\textbf{6}E-4 & 4.687144\textbf{4381616444}E-4 &4.6871441035708\textbf{09}E-4\\
6   & 6.106620777037364E-4  & 6.1066207770373\textbf{54}E-4 & 6.10662\textbf{12129571923}E-4 &6.106620777037\textbf{133}E-4\\
7   & 8.116930153767311E-4  & 8.1169301537673\textbf{53}E-4 & 8.116930\textbf{7331926445}E-4 &8.116930153767\textbf{775}E-4\\
8   & 1.106000581484408E-3  & 1.1060005814844\textbf{15}E-3 & 1.106000\textbf{6604360253}E-3 &1.106000581484\textbf{175}E-3\\
9   & 1.564054962964779E-3  & 1.5640549629647\textbf{81}E-3 & 1.56405\textbf{50746145116}E-3 &1.564054962964\textbf{430}E-3\\
10  & 2.500274605795732E-3  & 2.50027460579573\textbf{6}E-3 & 2.500274\textbf{9088856207}E-3 &2.\textbf{496489446288155}E-3\\
11  & 3.428396737487712E-3  & 3.4283967374877\textbf{04}E-3 & 3.42839\textbf{72314397134}E-3 &3.42\textbf{0841483175205}E-3\\
12  & 3.861374139996904E-3  & 3.86137413999690\textbf{1}E-3 & 3.861374\textbf{6648569073}E-3 &3.8\textbf{53287806877850}E-3\\
13  & 4.111139415564093E-3  & 4.11113941556409\textbf{7}E-3 & 4.111139\textbf{9582535370}E-3 &4.1\textbf{02938786513448}E-3\\
14  & 4.243541894555503E-3  & 4.24354189455550\textbf{5}E-3 & 4.24354\textbf{24466964718}E-3 &4.2\textbf{35268733626918}E-3\\
15  & 4.285156168585589E-3  & 4.2851561685855\textbf{95}E-3 & 4.285156\textbf{7236971821}E-3 &4.2\textbf{76333953447218}E-3\\
\hline
\end{tabular}
\label{t3}
\end{sidewaystable}

\clearpage

In order to test performances  of our closed-form solutions for varying   density contrasts, the same prismatic body in Figure \ref{testmodelprism} is tested, but with quartic order density contrasts. The density contrast is mixed in both horizontal and vertical directions:
\begin{eqnarray}
\lambda(\mathbf{r})=x^2yz,
\label{C1}
\end{eqnarray}
where the density is in units of $\,\text{kg/m}^3$ and $x, y, z$ are in units of $\text{km}$. Totally, a number of 256 observation sites are uniformly arranged at a plane with horizontal coordinates ranging from $0\,\text{km}$ to $30\,\text{km}$ in both $x$ and $y$ directions. The measuring plane has a vertical offset of $z=-2\,\text{km}$ to the top surface of the prism, which means there are no singularities in the gravity field vector. Due to no closed-form solutions available, we have to use the high-order Gaussian quadrature rule with $512\times512\times512$ quadrature points \citep{Davis1984Methods} to compute the reference solutions. Because the gravity field on this measuring plane is regular, therefore,  the high-order Gaussian quadrature could generate reliable reference solutions. 

\begin{figure}
\centering
\includegraphics[width=17cm]{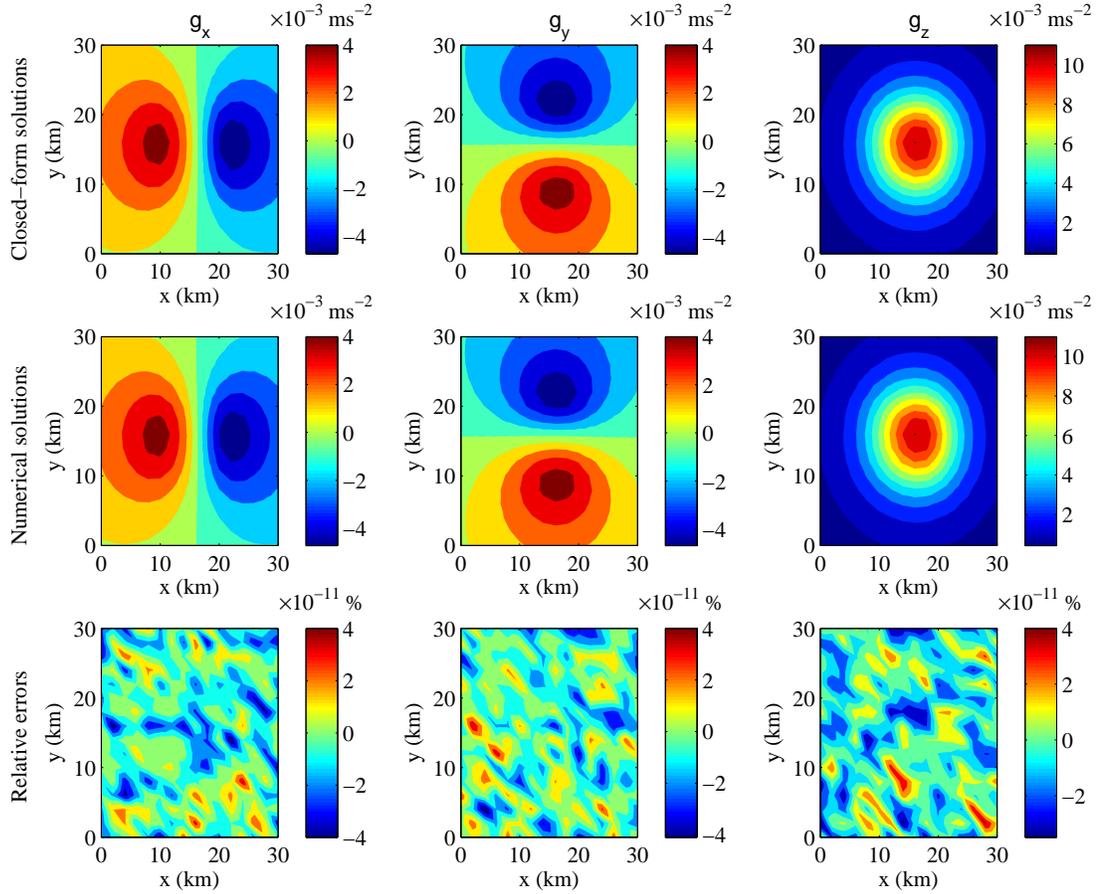}
\caption{Comparison of computed three components  of gravity fields ($g_x, g_y, g_z$) calculated by both our  formulae (upper row) and the high-order Gaussian numerical quadrature rule  with $512\times512\times512$ quadrature points (middle row)  for the case the measuring profile is above the prismatic body. The relative errors are shown in the bottom row. }
\label{prismx2yz}
\end{figure}

We compare the gravity fields computed by both our formula and the high-order Gaussian quadrature rule on this measuring plane, which are shown in Figure \ref{prismx2yz}. As expected, excellent agreements  are obtained for all the three components of the gravity fields. The relative errors between our solution and the high-order Gaussian quadrature rule's solution are less than $10^{-10}\%$. As presented in Table \ref{t4}, the maximum absolute residual of the computed gravity field is $1.969\times10^{-12}\,\text{ms}^{-2}$, which is far less than the  instrument precision of typical gravimeters used in current gravity surveys  (such as approximately $5\cdot10^{-8}\,\text{ms}^{-2}$ for CG-5 gravimeter \citep{Reudink2014}). The mean square  residuals  are less than $3.507\times10^{-14}\,\text{ms}^{-2}$. These almost negligible residuals verify the high accuracy of our new formula for the case of varying density contrasts.  However, happening to almost all analytical formulae  \citep{Holstein2003,Ren2018SG,jiang2018sg,chen2018gji}, phenomenon of numerical instability still exist in our new  formula when the distance between the observation site and the polyhedral mass body beyond a critical level. This phenomenon is caused by the limited machine precision to present the real number in the calculation, which can be resolved by  using longer bits such as 128 bits to represent real numbers or  equivalent real number representation methods.

\clearpage

\begin{table}
\centering
\caption{Residuals statistics between our solutions and those from the high-order Gaussian quadrature rule with $512\times512\times512$ quadrature points for gravity fields shown in Figure \ref{prismx2yz}.}
\begin{tabular}{crrr}
\hline
    & $g_x (\,\text{ms}^{-2})$ &  $g_y (\,\text{ms}^{-2})$ &  $g_z (\,\text{ms}^{-2})$\\\hline
Max    & $1.890\times10^{-12}$  & $1.069\times10^{-12}$  &$1.969\times10^{-12}$\\
Min    & $-1.230\times10^{-12}$  & $-9.996\times10^{-13}$  &$-2.903\times10^{-12}$\\
Mean   & $2.005\times10^{-14}$  & $1.881\times10^{-14}$  &$3.507\times10^{-14}$\\
\hline
\end{tabular}
\label{t4}
\end{table}

\clearpage

\section{Discussion and Conclusions}
We report the existence of closed-form solutions of gravity field for a  polyhedral body with general density contrasts. The density contrast is represented by a  polynomial function of arbitrary orders.  This polynomial density function can vary in both horizontal and vertical directions. Our closed-form solutions are singularity-free which means that the observation sites can be located at any place outside, inside, on the vertices of and at the edges of the polyhedral bodies. A synthetic prismatic body with different density contrasts are tested to verify our formula's accuracies. Excellent agreements between our new solutions and other solutions verify the capability of our new findings to accurately calculate gravity fields. With our new findings, the door of room containing closed-form solutions for polyhedral bodies with polynomial density functions can be closed.

\acknowledgments
This work was financially supported by the National Science Foundation of China (41574120), a joint China-Sweden mobility project (STINT-NSFC, 4171101400), and the China Scholarship Council Foundation (201806370223), the National Basic Research Program of China (973-2015CB060200).

\bibliography{mybib}

\end{document}